\documentclass{optica-article}
\journal{opticajournal} 
\articletype{Research Article}
\usepackage{lineno}

\usepackage{graphicx}  
\usepackage{dcolumn}   
\usepackage{bm}        

\usepackage{xspace}
\usepackage{amsmath}
\usepackage{abbrevs}
\usepackage[]{units}	
\usepackage{color}

\usepackage{float} 
\usepackage{wrapfig} 
\usepackage{multirow} 
\usepackage{rotating} 

\begin{document}

\title{Storage of telecom wavelength heralded single photons in a fiber cavity quantum memory}

\author{K.~A.~G. Bonsma-Fisher,\authormark{1,2} 
R. Tannous,\authormark{1}
D. Poitras,\authormark{1}
C. Hnatovsky,\authormark{1} 
S.~J. Mihailov,\authormark{1}
P.~J. Bustard,\authormark{1,*}
D.~G. England,\authormark{1,$\dagger$}
and B.~J. Sussman\authormark{1,2}}

\address{\authormark{1}National Research Council of Canada, 100 Sussex Drive, Ottawa, Ontario K1A 0R6, Canada\\
\authormark{2}Department of Physics, University of Ottawa, Advanced Research Complex, 25 Templeton Street, Ottawa, Ontario K1N 6N5, Canada}

\email{\authormark{*}philip.bustard@nrc-cnrc.gc.ca,
\authormark{$\dagger$}duncan.england@nrc-cnrc.gc.ca, http://quantumtechnology.ca}


\begin{abstract}
We demonstrate the storage and retrieval of heralded single photons in a fiber-based cavity quantum memory. 
The photons are stored, and retrieved, from the memory using quantum frequency conversion which switches the photon into, and out of, resonance with the cavity. The photons, generated in the telecom O-band with a bandwidth of 81\,GHz, are retrieved from the memory with a $1/e$ lifetime of 1.64\,$\mu$s, or 32.8 cavity round trips. We show that non-classical photon statistics remain for 70 round trips. The internal memory efficiency after 0.5\,$\mu$s of storage is $10.9 \pm 0.5$\%; a coupling efficiency of 60\% into the memory cavity yields a total efficiency of  $6.0\pm0.3$\%. These results mark a crucial step forward in the development of fiber-based quantum memories, and high-bandwidth memories operating at telecom wavelengths, with applications to photon source multiplexing and fiber-based quantum networking. 
\end{abstract}

Quantum memories are a vital component~\cite{Heshami2016} for a number of emerging quantum technologies including cryptography~\cite{BB84}, communication~\cite{Duan2001} and computing~\cite{KLM}. For long-distance quantum communication, memory lifetimes on the ms-scale are required, which can been achieved by mapping narrow-band light onto long-lived material excitations~\cite{Cho2016}. At the other extreme, local quantum processing tasks, such as photon synchronisation~\cite{Nunn2013}, require memories with high time-bandwidth products, enabling many independent operations within the memory lifetime. In this regime, the requirements on storage time are less stringent and the storage bandwidth becomes an important metric.

Time-bandwidth products in excess of $10^4$ have been realized using the atomic frequency comb (AFC) protocol~\cite{Wei2022}, and $\sim 1$\,GHz bandwidths have been achieved using the off-resonant cascaded absorption (ORCA)~\cite{Kaczmarek2018} and fast ladder memory (FLAME) protocols~\cite{Finkelstein2018}. While THz-bandwidth storage has been achieved using the Raman storage protocol~\cite{Bustard2013, England2015}, storage times were prohibitively short for local processing, and high-bandwidth storage of quantum light in material excitations remains an open challenge. A promising alternative technique is to store photons in free-space optical cavities with active switching~\cite{Pittman2002}. As the photons do not need to be bandwidth-matched to a material resonance, optical cavities can be used to store high bandwidths for many cavity cycles~\cite{Yoshikawa2013,Kaneda2015,Bouillard2019}. While the lifetime is insufficient for quantum communications, these cavity memories hold great promise for quantum processing applications with photon synchronisation~\cite{Kaneda2017}, high efficiency single photon generation~\cite{Kaneda2019}, and multiphoton state generation~\cite{Meyer-Scott2022} already demonstrated.

Optical fiber can provide low-loss long-distance propagation and compact local networks in a single robust platform. As such, fiber optics will be essential to developing future scalable quantum photonic technologies. In this direction, advances have been made in the development of fiber-integrated quantum components including quantum light sources~\cite{Fiorentino2002,Liu2022} and detectors~\cite{Hopker2019}, all-optical switches operating at the single photon level\cite{Hall2011,England2021}, and fiber-integrated versions of traditional quantum memory systems~\cite{Saglamyurek2015,Sprague2014}. Recently we have developed a quantum memory protocol known as fiber cavity storage with intra-cavity frequency translation (FC-SWIFT)~\cite{Bustard2022,Bonsma-Fisher2023} which is complementary to existing free-space~\cite{Pittman2002,Yoshikawa2013,Kaneda2015,Kaneda2017,Bouillard2019,Kaneda2019,Meyer-Scott2022} and in-fiber~\cite{Lee2023} cavity memories based on active optical switching. FC-SWIFT employs a monolithic all-fiber cavity with no active optical elements; instead storage and retrieval from the memory is driven by nonlinear interactions that change the frequency of the light. As the cavity is formed from a single fiber, it is inherently low loss offering more cavity cycles than would be possible in a fiber cavity containing active optical elements. Furthermore, since the photons are stored in fiber, they are immediately mode-matched with other elements in a network. In this work, we demonstrate an important milestone in the FC-SWIFT platform by storing and retrieving telecom O-band single photons produced by a spontaneous parametric down-conversion (SPDC) source. We show that single photons retrieved from the cavity retain non-classical statistics for up to 3.5\,$\mu$s of storage time, or 70 round trips.

\begin{figure}
\center{\includegraphics[width=0.6\columnwidth]{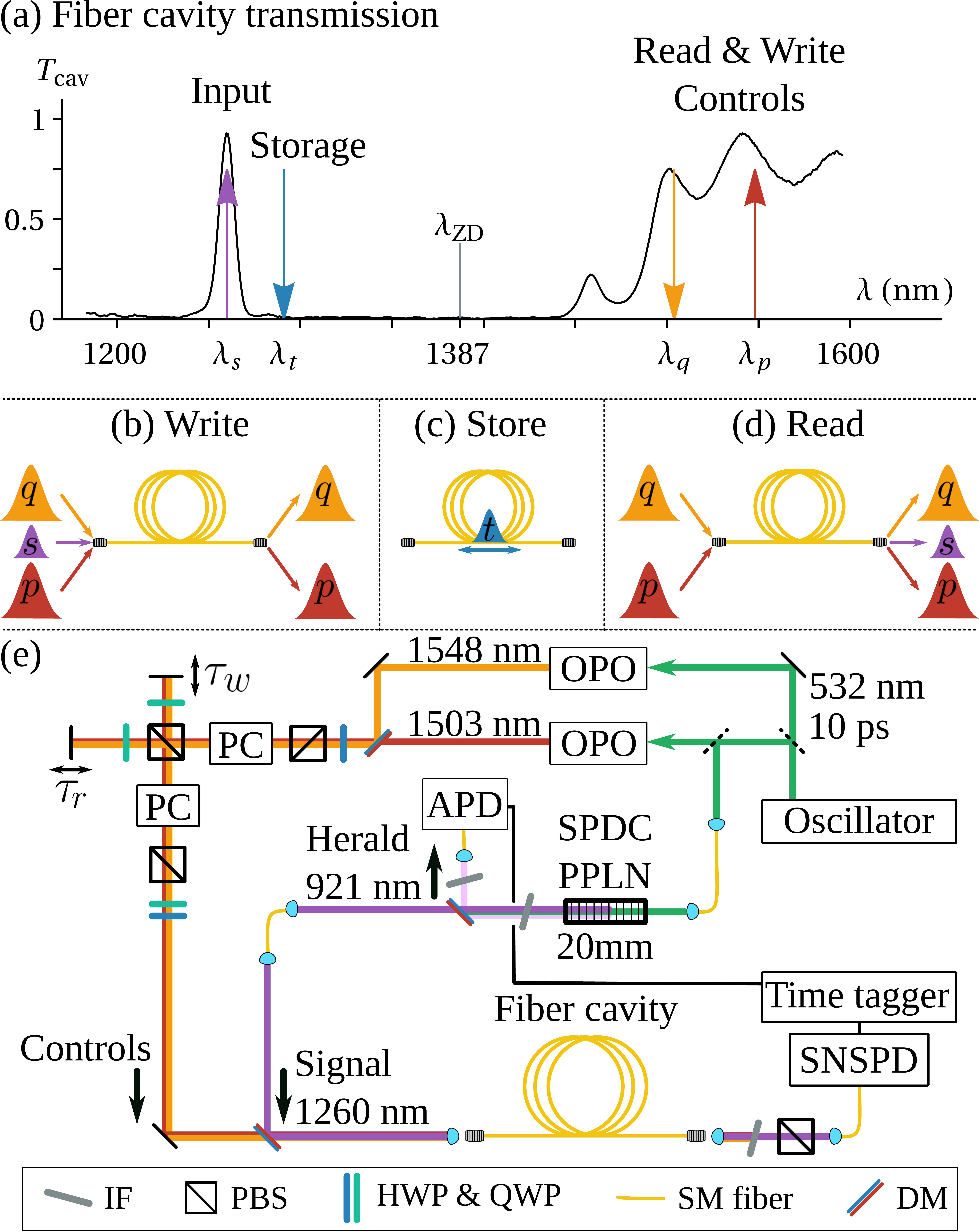}}
\caption{ 
(a) Fiber cavity transmission spectrum and zero dispersion wavelength.
Wavelengths $\lambda_s$, $\lambda_p$, and $\lambda_q$ are highly transmitted through the coating, while $\lambda_t$ is highly reflected. 
Fields $p$ and $q$ are intense pulses which drive BSFWM, translating $\lambda_s \to \lambda_t$. The zero-dispersion wavelength of the fiber is $\lambda_{ZD} \simeq 1387$\,nm. 
(b) The input signal $s$ enters the fiber cavity with the $p$ and $q$ write control pulses. 
(c) The signal is mapped to the cavity stored mode $t$ which reflects off the coating, circulating the cavity, until 
(d) The $p$ and $q$ read control pulses drive the reverse process and the signal exits the cavity in the $s$ mode. 
(e) Experimental setup.
IF: interference filter; PBS: polarizing beam splitter; HWP \& QWP: half- and quarter-wave plate; SM: single-mode; DM: dichroic mirror; PC: Pockels cell.
}\label{fig:setup}
\end{figure}

 We prepare our fiber-cavity quantum memory by depositing dichroic coatings (see Fig.~\ref{fig:setup}(a)) on the end facets of a commercially available single mode fiber.
The signal pulse $s$ propagates into the fiber cavity with high transmission, along with intense write control pulses, labelled $p$ and $q$. The write control pulses drive Bragg scattering four-wave mixing (BSFWM), so that the signal photon frequency is translated from $\omega_s$ to $\omega_t$. While the write control pulses exit the fiber, transmitting through the exit facet coating, the translated photon $t$ reflects off the coating and is trapped in the fiber cavity (see Fig.~\ref{fig:setup}(c)). After a pre-determined number of round trips, a pair of ($p,q$) read control pulses enter the cavity and drive the reverse frequency translation by BSFWM.
The signal photon is mapped back to its original frequency and exits the cavity through the exit facet coating with high transmission probability (see Fig.~\ref{fig:setup}(d)).

\begin{figure}
\center{\includegraphics[width=1.\columnwidth]{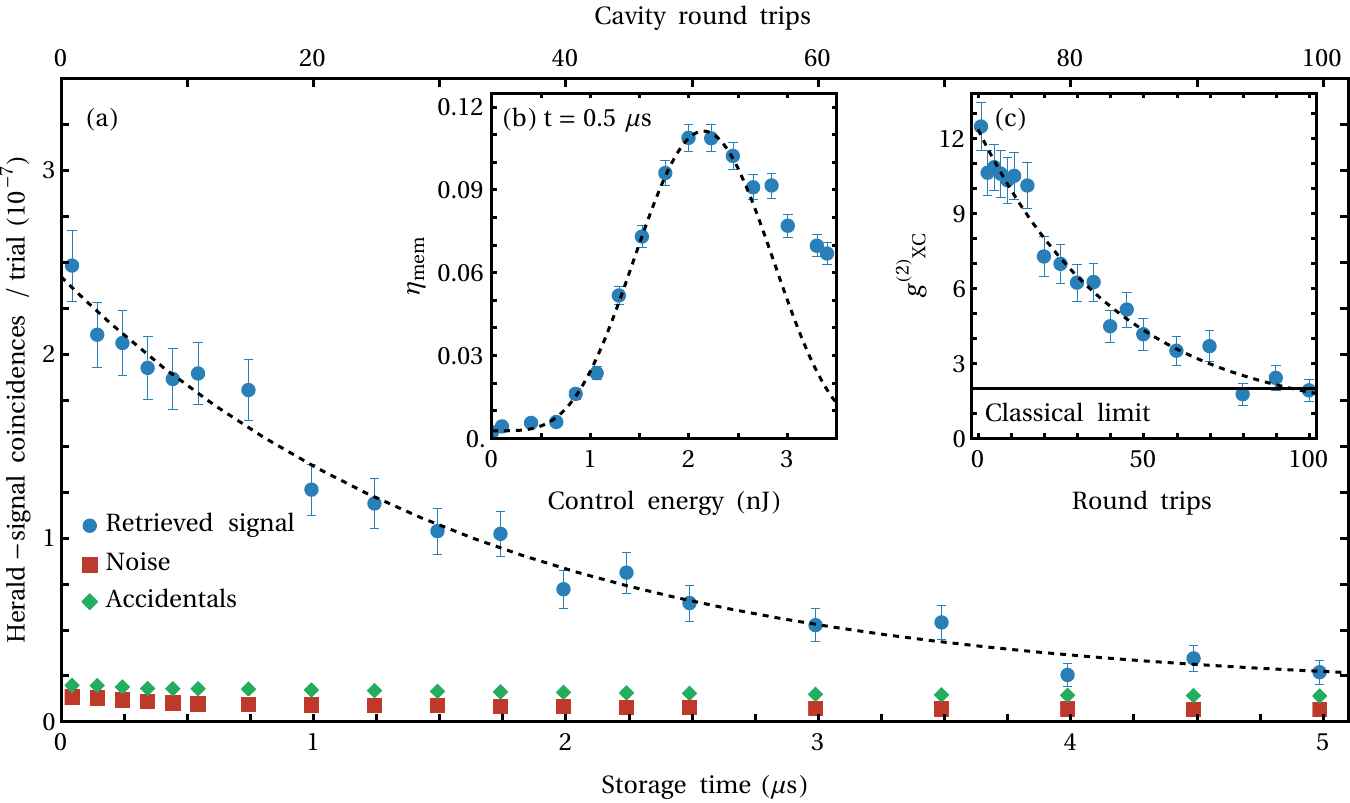}}
\caption{ 
(a) Retrieved signal photons (blue), noise (red) and accidentals (green) over storage time. 
Black dashed line is an exponential decay fit the data, with a $1/e$ decay constant of 1.64\,$\mu$s, or 32.8 round trips. 
(b) Memory efficiency  with total (read \& write) control pulse energy after 10 round trips ($t = 0.5$\,$\mu$s) of storage.  
The memory efficiency peaks at $10.9 \pm 0.5 \%$  for 2\,nJ of total control pulse energy. 
The black dashed line is a $\sin^4 (P)$ fit to the data up to 2.5\,nJ, where $P$ is the control energy.
(c) The retrieved signal remains non-classical, i.e., $g^{(2)}_{XC} > 2$, for 70 round trips (3.5\,$\mu$s) of storage.
(a-c) Uncertainties are calculated assuming Poissonian error in photon detection. 
}\label{fig:readout}
\end{figure}

The experimental setup is shown in Fig.~\ref{fig:setup}(e).
Pulses from the primary laser (80.1\,MHz oscillator, 532\,nm 10\,ps pulses) pump two optical parametric oscillators (OPOs) to generate the $p$ and $q$ control pulses at wavelengths $\lambda_p = 1548$\,nm and $\lambda_q = 1503$\,nm, respectively.
Control pulses $p$ and $q$ are overlapped on a dichroic beam splitter and pulse picked in the \emph{write} and \emph{read} time-bins by Pockels cells (PCs), with fine adjustment of write and read control pulse delays set by optical delay lines.
A portion of the 532\,nm primary laser beam pumps SPDC in a periodically-poled lithium niobate (PPLN) crystal to produce signal and herald photon pairs at $\lambda_s=1260$\,nm and $\lambda_h=921$\,nm, respectively. The signal photon is overlapped with the write control pulses on a dichroic beam splitter and coupled into the memory cavity with coupling efficiency $\eta_\text{in} = 0.55$.
The Pockels cells are operated at a rate of 181\,kHz, allowing storage times up to 5.5\,$\mu$s. This ensures that any light generated in the cavity from the previous storage has escaped.
The detection of a herald photon on an avalanche photodiode (APD) heralds the presence of a signal inside the fiber cavity with 9.1\% efficiency. Signal photon losses include multiple mirror reflections, spectral filtering, a dichroic mirror to combine with the pumps, and coupling into the memory cavity: all of which contribute to a modest heralding efficiency. Improvements in heralding efficiency would directly improve the signal-to-noise ratio of this demonstration.  
 
The cavity fiber is a $\sim 5$\,m-long polarization-maintaining, highly-nonlinear fiber (\emph{PMHN5}), with a cavity cycle time of $\tau_\text{cav} = 49.877$\,ns. The cavity cycle time can be tuned by mechanically stretching the fiber. In this way it is possible to bring the cavity into resonance with the laser repetition rate ($4\tau_\text{laser} = 49.938$\,ns). However, due to slipping and hysteresis in the current stretching mechanism, it was not possible to perfectly match the fiber cavity and the laser cycle times, leaving a residual mismatch of $\tau_r = 2.6$\,ps. This mismatch would reduce the effective memory lifetime to just $\simeq5$ cavity round trips. However, in this demonstration we correct for this offset by physically changing the control pulse delay time to achieve storage for over 70 round trips. In future applications, where the required storage time cannot be known in advance, this timing mismatch must be resolved.

The dielectric coatings applied to each fiber end-facet show $> 90\%$ transmission at $\lambda_s=1260$\,nm and high reflectivity at $\lambda_t = 1291.5$\,nm. A loss per round trip of 0.08\,dB was extracted from a cavity ring-down measurement~\cite{Bustard2022} using bright pulses at $\lambda_t$. The loss of the fiber is quoted to be $<0.9$\,dB/km ($<0.009$\,dB per round trip), so we expect that, in this configuration, the memory lifetime is limited by reflection losses at the cavity mirrors. For longer cavities, the fiber loss could support memory lifetimes as long as 50\,$\mu$s, though longer round-trip times may become impractical for on-demand retrieval. The signal, write, and read polarizations are all aligned with the slow axis of the fiber. 

The frequency splitting between the $p$ and $q$ control pulses is chosen to match the desired separation of the input and stored signal frequencies, such that $(\omega_q-\omega_p)=(\omega_s-\omega_t)$, where $\omega_i = 2 \pi c / \lambda_i$. Phase matching of BSFWM requires that $\Delta \beta = \beta(\omega_p)  - \beta(\omega_q) + \beta(\omega_s) - \beta(\omega_t) = 0$, where $\beta(\omega)$ is the wave vector. Here, this is achieved with the signal and control fields on opposite sides of the fiber zero-dispersion wavelength, $\lambda_\text{ZD} \sim 1387$\,nm, such that $p$ and $s$ are group-velocity matched, as are $q$ and $t$. Write control pulses are picked by the PCs to match the input at $t=0$\,$\mu$s, and read control pulses are picked to overlap with the stored photon after integer $\mathcal{N}$ round trips at $t = \mathcal{N}\times( 4 \tau_\text{laser} + \tau_r)$. On exiting the fiber cavity, the control pulses are removed by short-pass spectral filters and a 1.3\,nm band-pass around 1260\,nm is used to suppress background noise generated by competing nonlinear processes. The signal photon is polarization filtered for maximum transmission, collected into a single-mode fiber, and directed to a superconducting nanowire single-photon detector (SNSPD in Fig.~\ref{fig:setup}).  
Coincidence measurements are made between signal and herald detections, predicated on the Pockels cell operation.

\begin{figure}
\center{\includegraphics[width=0.5\columnwidth]{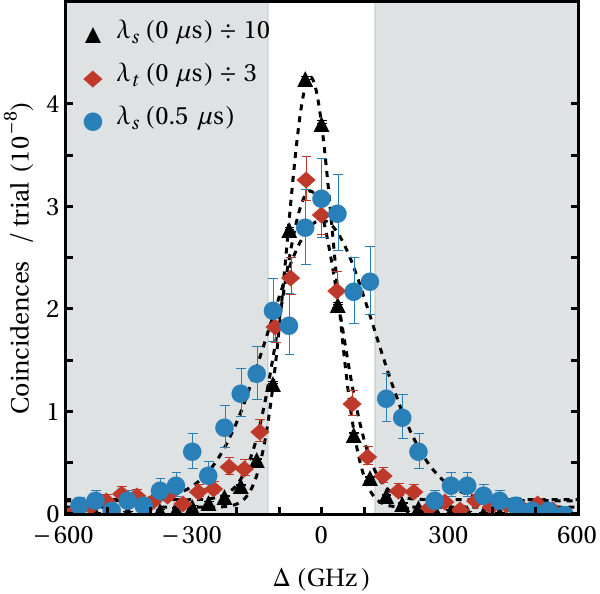}}
\caption{
Monochromator measurement of the input signal (black triangles), and the retrieved signal after 10 round trips (filled blue circles), both measured at $\lambda_s = 1260$\,nm. The stored light at $\lambda_t = 1291.5$\,nm cannot be measured due to the high reflectivity of the cavity. Instead, a reference fiber of the same length but with no reflective coating is used to measure the spectra of the stored light (red diamonds).
Gaussian fits to the data (black dashed lines) give FWHMs of 81\,GHz, 130\,GHz, and 275\,GHz for the input, stored, and retrieved signals, respectively, where the monochromator resolution (100\,GHz, or 0.53\,nm) has been deconvolved.The input and stored spectra are divided by 10 and 3 respectively for comparison. Uncertainties are calculated assuming Poissonian error in photon detection. Gray shading shows edges of the spectral filtering used for all other measurements. }
\label{fig:spec}
\end{figure}

In Fig.~\ref{fig:readout} we plot the herald-signal coincidence counts per trial as a function of write-read delay, or storage time, up to 5\,$\mu$s. An exponential decay curve is fit to the data giving a decay constant of 1.64\,$\mu$s, or 32.8 round trips. The discrepancy between this decay and the cavity ring down measurement with bright pulses is due to fiber dispersion increasing the duration of the stored photon as it propagates. This reduces the read efficiency at longer storage times, and shortens the memory lifetime.

Noise counts, produced when only control pulses are directed to the cavity, are plotted alongside the data. Noise photons are produced in the output signal bin by anti-Stokes Raman scattering of the control pulses at a rate of $3.5 \times 10^{-4}$ photons per shot. We also plot the accidental signal-herald coincidences, which are uncorrelated detection events coinciding in time.
Since we are pumping the SPDC source at 80.1\,MHz, signal photons produced in the read time-bin can coincide with a herald photon produced at $t=0$\,$\mu$s with probability $P_\text{acc} = P_s P_h = N_s N_h / R^2$, where $N_{s(h)}$ is the signal (herald) singles rate, and $R$ is the repetition rate. 
As seen in Fig.~\ref{fig:readout}(a), the memory operation is primarily limited by accidental coincidences, rather than noise photons; we note that using an additional PC to gate the SPDC pump would largely eliminate these accidental coincidences. 
We compute the cross-correlation second-order coherence function $g^{(2)}_\text{XC} = P_{s,h} / P_s P_h$, plotted as a function of storage time in Fig.~\ref{fig:readout}(c).
The measured $g^{(2)}_\text{XC}$ remains above the classical limit of 2~\cite{Clauser1974} up to 70 round trips (3.5\,$\mu$s) of storage. 
For the input photon, at $t=0\,\mu$s, we measure $g^{(2)}_\text{XC} = 180$.

Figure~\ref{fig:readout}(b) shows the internal memory efficiency  $\eta_\text{mem} = N_{s,h}(t) / N_{s,h}(0)$ at retrieval delay $t=0.5\,\mu$s, as a function of total control pulse energy.
We observe a maximum memory efficiency of $10.9\pm0.5\%$ for 2\,nJ of total control pulse energy, evenly split between write and read, and $p$ and $q$, fields.  
The efficiency of frequency translation by BSFWM~\cite{McKinstrie2002, Lefrancois2015} is given by $\eta_{\text{BSFWM}} = (4\gamma^2 P^2 / \kappa^2) \sin^2 \kappa L$, where the fiber has length $L$ and nonlinear coefficient $\gamma$, $P^2 = P_p P_q$ is the control power, and $\kappa = \sqrt{(\Delta \beta/2)^2 + 4\gamma^2 P^2}$.
Since the memory operation requires two separate BSFWM processes, the internal memory efficiency scales $\propto \sin^4 2 \gamma PL$ when $\Delta \beta \sim 0$.
At higher control pulse energies, $\eta_\text{mem}$ deviates from the $\sin^4 2 \gamma PL$ fit, due to competing nonlinear processes including self- and cross-phase modulation. We note that the total memory efficiency, including coupling efficiency $\eta_\text{in}$, after $t=0.5\,\mu$s of storage is $\eta_\text{tot} = \eta_\text{in} \eta_\text{mem} = 6.0\pm0.3\%$.

\begin{figure}
\center{\includegraphics[width=0.5\columnwidth]{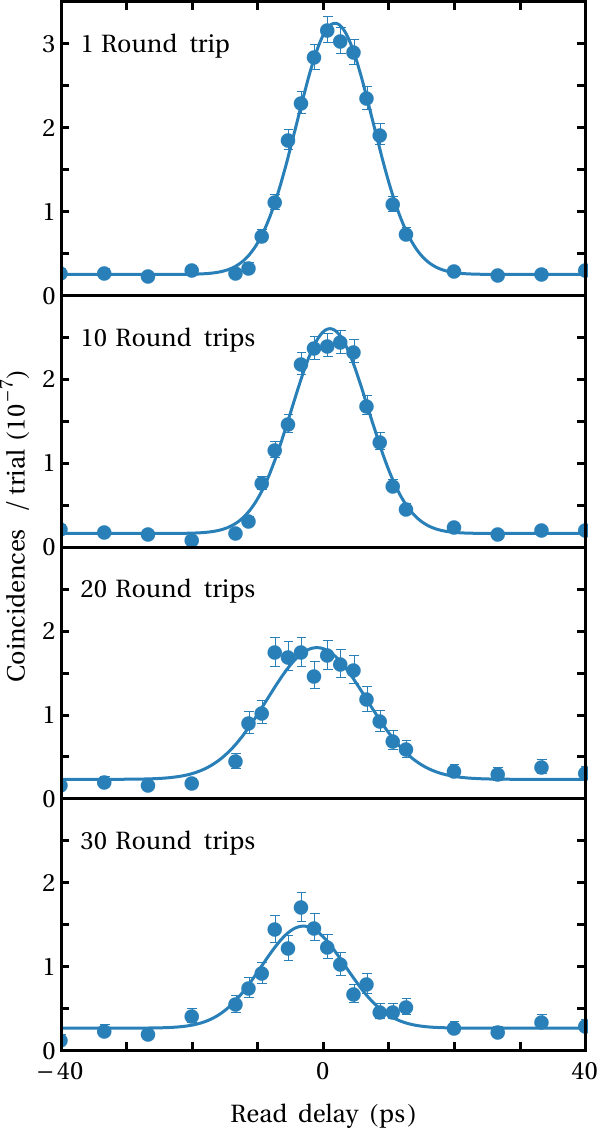}}
\caption{The read control delay is scanned relative to the heralded single photons  for 1, 10, 20, and 30 round trips (RT). Data are fit by Gaussians of FWHM $13.6\pm 0.3$\,ps, $13.8\pm 0.4$\,ps, $18\pm 1$\,ps and $15\pm 1$\,ps respectively. The shape and duration of the curves do not significantly change with increasing storage time, implying that dispersion in the memory is minimal on these timescales. See text for details.}
\label{fig:fwhm}
\end{figure}

Next, we measure the spectra of the heralded input photons, and the retrieved photons, by directing photons collected from the memory to a scanning monochromator (resolution 100\,GHz) before the SNSPDs (see Fig.~\ref{fig:spec}). As we cannot directly access the stored photons due to the reflective coatings, we use a \emph{reference fiber} of the same length, but without coatings, to measure bandwidth of the stored photons. Gaussians fitted to the input, stored, and retrieved spectra give FWHMs of 81\,GHz, 130\,GHz, and 275\,GHz, respectively, after deconvolving the monochromator resolution. These measurements indicate that each action of the control pulses broadens the spectrum of the photons; we attribute this broadening to cross-phase modulation (XPM) as has been observed in other BSFWM experiments~\cite{Bonsma-Fisher2022}. Such broadening reduces the fidelity of the storage process and, when combined with tight spectral filtering, is a source of loss. The transmission of the spectral filter used for all other measurements is plotted in shaded gray in Fig.~\ref{fig:spec}. Due to spectral broadening, 30\% of the retrieved photons are blocked by this filter, representing a significant reduction in the total memory efficiency.

In this demonstration, the signal and control pulses are of comparable duration. This means that the control intensity is not uniform across the signal pulse, and it is impossible to efficiently convert the entire signal pulse. Additionally, the effects of pulse walk-off due to mismatched group velocities further reduce the conversion efficiency. These factors, along with the broadening shown in Fig.~\ref{fig:spec}, result in a reduced memory efficiency. In future experiments, longer control pulses could be used to increase the memory efficiency, reduce the effects of XPM, and improve the overlap between input and output spectra. In the limit where the control pulses are much longer than the signal and the walk-off time, unit conversion efficiency can, in principle, be achieved. However, practical limitations such as laser power and damage thresholds may preclude this. Alternatively, use of control-signal pulse walk-through by a crossed-polarization configuration in PM-fiber can equalize the interaction strength and permit unit efficiency~\cite{Christensen2018}.

We also investigate the temporal profile of the retrieved signal by performing fine resolution read control pulse delay scans for a selection of storage times. Figure~\ref{fig:fwhm} shows the read delay scans at 1, 10, 20, and 30 round trips of storage. Gaussian fits to the data show FWHMs of $13.6\pm 0.3$\,ps, $13.8\pm 0.4$\,ps, $18\pm 1$\,ps, and $15\pm 1$\,ps respectively, where the quoted uncertainty is due to statistical fluctuations, and does not reflect experimental drift between measurements. In previous FC-SWIFT demonstrations, second and third-order dispersion from the reflective elements has significantly distorted the temporal profile of the stored photons~\cite{Bonsma-Fisher2023}. Here, there is no evidence of third-order dispersion, and the minimal broadening is consistent with second-order dispersion only from propagation in the fiber: 0.31\,ps$^2$ per cavity cycle as measured using white light interferometry.

In conclusion, we have demonstrated storage of non-classical light in a fiber cavity memory with a $1/e$ lifetime of 1.64$\mu$s or 32.8 round trips. The efficiency, including coupling losses, is $6.0\pm0.3$\%. An internal memory efficiency of $\sim 10\%$ was measured, which we expect can be improved in future experiments by modifying the frequency conversion process. In particular, since the signal and control pulses are pair-wise group-velocity matched, we require the control pulse durations to be significantly longer than the signal to achieve high efficiencies. We found that the memory was not limited by noise, but rather by accidental coincidence events from the photon source, which can be extinguished with pulse picking. We observed a cavity ring-down of 32.8 round trips ($1/e$) which determines the number of independent operations that can be performed on a stored pulse during the memory lifetime. However, it is worth noting that this storage time is over $10^5$ times longer than the input pulse duration. Accordingly, several thousand time bins could be stored simultaneously allowing parallel memory operations.
Operating at other signal wavelengths, we measured cavity lifetimes in excess of 100 round trips, albeit with reduced signal-to-noise operation. We expect that future coating designs can achieve longer-lived storage with the same low-noise properties seen here. 
In contrast to other fiber-based memory schemes based on active switching~\cite{Hoggarth2017, Lee2023}, the use of a monolithic fiber with no active elements reduces the cavity loss to less than 1\% per round trip: comparable with state-of-the-art free-space cavity memories~\cite{Kaneda2019}, with the potential for further improvement with future cavity designs. However, unlike memories based on active switching which have near-unit efficiency, the memory efficiency is $<10$\%. Improving this efficiency will be the focus of future work. Finally, due to the low-loss propagation of telecom light in optical fiber, much longer cavities are feasible. For example, in the current fiber, a 500\,ns cavity cycle time could be implemented with minimal performance degradation; such length scales are challenging in free-space cavities. With these suggested improvements, we anticipate that the the FC SWIFT memory will be of use in local quantum processing tasks, such as multiplexing probabilistic photon pair sources, in the near future.

\begin{backmatter}

\bmsection{Acknowledgments}
We are grateful for discussions with Khabat Heshami, Aaron Goldberg, Fr\'{e}d\'{e}ric Bouchard, Kate Fenwick, Guillaume Thekkadath and Yingwen Zhang. We also thank Denis Guay, Rune Lausten, Rob Walker and Doug Moffatt for technical assistance. 

\bmsection{Funding}
This work was supported by the Quantum Research and Development Initiative (QRDI), led by the National Research Council, under the National Quantum Strategy.

\bmsection{Disclosures} 
The authors declare no conflicts of interest.

\end{backmatter}

\bibliography{FCSWIFT1260}
\end{document}